# Full Silicon Pillar-based 1D Optomechanical cavities

Juliana Jaramillo-Fernandez[a,b,†], Martin Poblet[a,b], David Alonso-Tomás[a,b], Christian Vinther Bertelsen[c], Elena López-Aymerich[d], Daniel Arenas-Ortega[a,b], Winnie E. Svendsen[c], Néstor E. Capuj[e,f], Albert Romano-Rodríguez[a,b,*], Daniel Navarro-Urrios[a,b,*]

[a] Departament d'Enginyeria Electrònica i Biomèdica, Universitat de Barcelona, 08028, Barcelona, Spain

[b] Institute of Nanoscience and Nanotechnology (IN2UB), Universitat de Barcelona, 08028, Barcelona, Spain

[c] DTU Bioengineering, Danmarks Tekniske Universitet (DTU), 2800 Kgs. Lyngby, Denmark

[d] DTU Nanolab, Danmarks Tekniske Universitet (DTU), 2800 Kgs. Lyngby, Denmark

[e] Depto. Física, Universidad de La Laguna, 38200 San Cristóbal de La Laguna, Spain

[f] Instituto Universitario de Materiales y Nanotecnología, Universidad de La Laguna, 38071 Santa Cruz de Tenerife, Spain

[†] Present address: Departament de Màquines i Motors Tèrmics, Universitat Politècnica de Catalunya, 08028, Barcelona, Spain

*Corresponding authors: albert.romano@ub.edu, dnavarro@ub.edu

Abstract

Nanomechanical resonators can serve as ultrasensitive, miniaturized force probes. While vertical structures like nanopillars are ideal for this purpose, transducing their motion is challenging. Pillar-based photonic crystals (PhCs) offer a potential solution by integrating optical transduction within the pillars. However, achieving high-quality PhCs is hindered by inefficient vertical light confinement. Here, we present a full-silicon 1D photonic crystal cavity based on nanopillars as a new platform with great potential for applications in force sensing and biosensing areas. Its unit cell consists of a silicon pillar with larger diameter at its top portion than at the bottom, which allows vertical light confinement and an energy bandgap in the near infrared range for transverse-magnetic (TM) polarization. We experimentally demonstrate optical cavities with Q-factors exceeding $10^3$ constructed by inserting a defect within a periodic arrangement of this type of pillars. Given the fact that that each nanopillar naturally behaves as a nanomechanical cantilever, the fabricated geometries are excellent optomechanical (OM) photonic crystal cavities in which the mechanical motion of each nanopillar composing the cavity can be optically transduced. These novel geometries display enhanced mechanical properties, cost-effectiveness, integration possibilities, and scalability, and opens and new path in front of the widely used suspended Si beam OM cavities made on silicon-on-insulator.

Introduction

The concept of one-dimensional photonic crystal waveguides (1D-PhCW)[1] has been extensively developed during the last decades in integrated photonics to create lasers,[2,3] optical filters[4] or optical cavities,[5] among other components, with applications in standard telecommunications,[6] quantum information processing[7] or high-sensitivity force-sensing[8] and biosensing[9]. If properly designed, these geometries can display a photonic band-gap in the near-infrared telecom windows along their periodic direction and can confine light in the lateral and vertical dimensions by exploiting the principle of refractive index guiding.[1] In this last regard, the most common way to effectively achieve light confinement relies on introducing different dielectric materials so that there is a sufficiently large refractive index contrast between the material constituting the 1D-PhCW and the surrounding media. Essentially, there are two categories of 1D-PhCW able to operate in the near-infrared range: i) nanobeams, which usually comprise a ridge or corrugated waveguide of a high refractive index material with air-holes and ii) nanopillars, also known as rods, which generally consist of a row of pillars of a high refractive index material separated by air gaps. The exploration of the latter category has been comparatively limited, largely due to the inherent challenges in fabricating vertical nanopillars that do not naturally bend and in achieving sufficiently large refractive index contrast in the vertical dimension to avoid light leakage. Indeed, an obvious advantage of the nanobeams is that they can be fabricated to be free-standing, i.e., surrounded by air, while nanopillars need a substrate to be anchored to. For these reasons, to the best of our knowledge, there is only a handful of experimental demonstrations of two-dimensional photonic crystal waveguides based on pillars[10–12] and none of a 1D-PhCW. Thus far, only numerical studies are documented in literature[13].

On the other hand, nanopillar mechanical resonators have garnered significant attention, particularly in biological applications, as extremely sensitive force sensors due to their outstanding mechanical properties[14–16]. However, their motion is particularly challenging to transduce, since most of the common methods used for horizontally oriented nanoelectromechanical systems, which usually rely on electrodes directly placed on top of the resonator[17], are not suitable. Thus, it is required high-resolution complex optical imaging techniques for their characterization[18–20].

In this manuscript we demonstrate the first 1D-PhCW geometrical configuration based on pillars that overcomes previous challenges and that can be fully fabricated on a single material wafer, e.g., monocrystalline silicon (Si) as in the present work. Indeed, we have successfully fabricated optical cavities by inserting an adiabatic defect within a 1D-PhCW made of Si pillars which display experimental optical quality factors exceeding $10^3$ at wavelengths around 1,5 μm. Moreover, since the pillars constituting the cavity region are at the same time a collection of mechanical cantilevers, we experimentally demonstrate that the fabricated 1D-PhCW pillar cavities are also excellent optomechanical (OM) photonic crystal cavities displaying large OM coupling rates ($g_{om}/2\pi$) exceeding 1MHz for mechanical modes oscillating at few tens of MHz. These geometries gather the outstanding mechanical properties of nanopillars within a high-quality photonic crystal cavity, paving the way for an alternative configuration of highly sensitive force sensors based on an optomechanical transduction mechanism, alternative to 1D-PhCW nanobeams.[21,22]

Results

The unit cell of the 1D-PhCW comprises monocrystalline silicon (Si) pillars with two vertical portions: a top Si portion with a height $t_1$ resting on top of the lower Si portion, with a height $t_2$. The radius of the top portion, denoted as r, is approximately $\Delta r=50$nm larger than that of the bottom portion (refer to inset of Figure 1a). The substrate is also composed of silicon, although employing materials like silicon oxide ($SiO_2$), with lower refractive index, would have negligible impact on the photonic behaviour for the geometry employed here. The unit cell is repeated with a pitch (a) along the propagation direction (x-axis). The photonic dispersion relation associated with the previous geometry, which is shown in the main panel of Figure 1a for the transverse magnetic TM-like polarized optical modes (blue lines), displays a bandgap for light with TM-like polarization and does not support TE-like modes. By choosing the right geometrical parameters (a=350nm, r=105nm and $t_1=t_2=1500$nm) the TM-like band gap can be positioned around 200 THz, with a gap-to-midgap ratio of 5.7%. The electric field distribution along the pillar axis for the lowest energy band at its band edge at $k_x=\pi/a$ (refer to inset of Figure 1b) is spatially localized in the top portion of the pillar, thus being isolated from the presence of the Si substrate. In the main panel of Figure 1b, we have plotted the energy of this optical mode as a function of a factor γ that rescales r and a, keeping the pillar height and $\Delta r$ unchanged. As expected, the band edge is pushed up in energy by decreasing γ, thus providing the needed intuition to create a defect within the 1D-PhCW that would place a resonant optical mode within its photonic band gap. Importantly, for heights of the lower portion of the pillar above a minimum value of $t_2\sim200$nm, no modification in the band diagram occurs if the rest of geometrical parameters are unchanged.

The 1D-PhCW pillar cavity is composed by two mirrors, each one of them consisting of 9 equivalent cells with the geometrical parameters associated to the band dispersion of Figure 1a. The defect region has been placed between the mirrors by inserting 10 or 11 central cells in which the pitch and the radius are progressively reduced from the outer cells in a quadratic way toward the centre to a minimum value of g*a and g*r, respectively. It is noteworthy that the energy of the fundamental optical cavity state can be tuned by varying the defect depth described by the factor g (see Figure 1d). A deeper defect (corresponding to a smaller scale factor g) results in higher energy of the optical cavity state. Here, we have also tested the possible influence of variations of $t_1$ and $t_2$ (see Supplementary Info), concluding that, above $t_2\sim200$nm, the optical quality factor (Q-factor) improves with $t_2$, while the resonance position remains unchanged. For $t_2<200$nm, radiative leaking towards the Si substrate prevents supporting any cavity mode. Importantly, for these latter geometrical parameters ($t_2<200$nm), the expected Q-factor is limited to $10^2$ due to radiative leaking even by employing materials with low refractive index like $SiO_2$ (see Supplementary Section S3). This limitation justifies the absence in the literature of experimental realizations of pillar-based 1D-PhCW geometries. On the other hand, for heights of the higher portion of the pillar above $t_1$~500nm the optical Q-factor increases while the energy of the mode decreases. For $t_1$<500nm the mode is cut-off, i.e., the mode leaks into the surrounding materials and attenuates. An illustration of the top and side view of a representative 1D-PhCW pillar cavity with g=0.75, a=350nm, r=105nm, $t_1$=1580nm and $t_2$=850nm is displayed in Figure 1d.

The 1D-PhCW pillar structures were fabricated by electron beam (e-beam) lithography and reactive ion etching (RIE) on an n-type <100> Si substrate. The wafer was prepared for e-beam exposure by spin-coating a 180 nm layer of CSAR positive e-beam resist. A 20 nm Aluminium layer was deposited using e-beam evaporation followed by a lift-off process to remove the resist, leaving the aluminium layer with the desired pattern as a mask for the subsequent dry etch. The RIE process was done in 2 steps to create the upper and lower portions of the nanopillar with different diameters. The first part of the etching process was done with a simultaneous mix of etching gas and passivation gas. For the second part of the process, the etch was changed to a cyclic process alternating between etching and passivation. The height of the nanopillars can be adjusted by changing the etching time in the first step and the number of cycles in the second step. Finally, the remaining aluminium mask was removed using a developer. Figure 1e shows a scanning electron micrograph (SEM) of one of the resulting geometries. More details on the fabrication recipes can be found in the Supplementary Section S2.

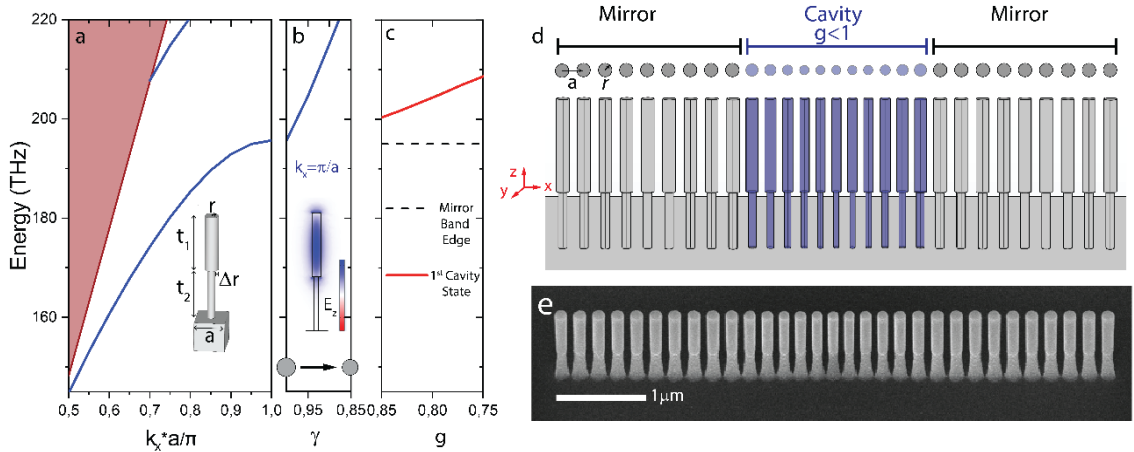

**Figure 1. Photonic properties of a one-dimensional photonic crystal waveguide (1D-PhCW) cavity composed of a linear array of nanopillars**. a) Photonic dispersion relation showing the TM-like polarized optical modes (blue lines) of an idealized mirror unit cell (depicted in the inset). The shaded region is the light cone. The 1D-PhCW has a bandgap for TM-like modes, centred around 200 THz. The unit cell has a lattice constant a=350 nm, pillar radius $r$=105 nm, $\Delta r$=50 nm, and top and bottom portions of the pillars have heights $t_1$=$t_2$=1500 nm. b) Dependence of the X-point band edge energy on the reduction factor γ. The spatial distribution of the electric field along the vertical direction of the pillar ($E_z$) is also illustrated for the state at the band-edge. c) Dependence of the energy of the optical cavity state on the depth of the defect region parametrized in terms of the reduction factor g, which determines the radius and pitch of the pillars

at the centre of the cavity. The optical cavity state is situated above the band-edge (illustrated with a dashed line). d) Schematic illustration of the geometrical parameters of the photonic crystal waveguide cavity as seen from the top and from the side. e) Scanning electron microscope tilted view (30 degrees) of a representative fabricated 1D-PhCW pillar cavity with scaling factor g=0.75.

In our experiments, the 1D-PhCW pillar cavity was replicated across multiple chips, varying the defect depth g by factors of 0.85, 0.8, and 0.75. Additionally, the entire 1D-PhCW cavity was scaled by a factor γ ranging from 1 to 1.1 in steps of 0.02. This allowed the mapping of the energy of the optical cavity modes and resulting variations in the mechanical response.

**Optical properties of the 1D-PhCW pillar cavity**

The optical properties of the samples were examined by evanescently coupling resonant laser light into the cavity. This was achieved using a fibre loop positioned above the cavity region, aiming to avoid physical contact with the pillars.fibreAn infrared driving laser with tunable wavelength (1355–1640 nm) and power up to 20 mW passes through a polarization controller before reaching the fibre loop. All the measurements were performed in an anti-vibration cage at ambient conditions. A schematic of the experimental setup, which allows both reflection and transmission configurations, is presented in Figure 2a and described in detail in the Supplementary Section S1. The 1D-PhCW pillar cavities are of the bi-directional type, i.e., light can decay in both forward and backward propagating modes of the fibre loop. Thus, it gives rise to an optical resonance that points upwards in a reflection spectrum when the input light excites the cavity via the evanescent field. The optical reflection spectrum of a representative fabricated 1D-PhCW pillar cavity with γ =1.06 and g=0.8 is shown in Figure 2d. By comparing the simulations and the experimental results it is possible to identify that the observed resonance corresponds to the fundamental optical cavity mode of the set of cavity modes supported by this geometry, which would in principle lead to a quality factor exceeding $10^4$ (see Supplementary Section S3). Figures 2b and 2c are, respectively, the top view and cross-section of the spatial profile of the simulated fundamental TM optical cavity mode above the band edge of a 1D-PhCW pillar cavity made with the idealized mirror cell as described in Figure 1c. For this cavity mode, the electric field is strongly localized in the cavity centre and decays rapidly in the mirror regions.

In the experiment, the optical resonance wavelength is centred around 1365.5 nm. The Lorentzian fit to the experimental data (refer to dashed curve of Figure 2d) allows estimating that the fabricated structure owns an overall optical quality factor of $Q=1.5 \times 10^3$. This value is very close to the intrinsic one considering that the coupled power fraction is rather low (<10%), resulting from a significant effective index mismatch between the propagating mode of the tapered fibre ($n_{eff}$~1.4) and the cavity state ($n_{eff}$~2.1). Intrinsic losses are, thus, dominating the overall Q-factor of the cavity, i.e., extrinsic losses due to the evanescent light coupling to the fibre can be neglected.

The intrinsic losses of the fabricated 1D-PhCW pillar cavities stem from scattering losses at the surface of the pillars, probably localized at the region where the supported mode overlaps with the undercut region, i.e., where the roughness of the pillar surface significantly increases.

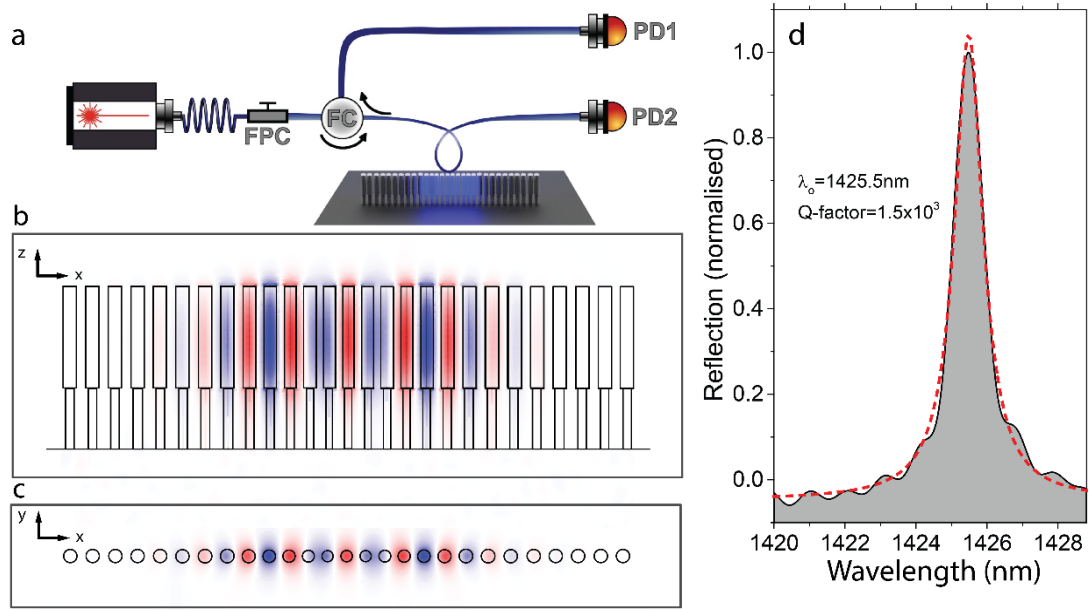

**Figure 2. Optical properties of the 1D-PhCW pillar cavity.** a) Schematic of the experimental setup. The fibre passes through a fibre polarization controller (FPC) and a fibre circulator (FC), enabling reflection and transmission experiments detecting with the photodetector (PD) 1 and 2, respectively. The fibre loop is positioned over the cavity region. Dimensions are not to scale. The diameter of the loop is about 50μm while the total length of the 1D-PhCW pillar cavity is about 10μm. b) and c) Finite-Element-Method simulation of the electric field along the z direction (Ez) of the TM fundamental optical cavity mode as seen from the side and from the top, respectively. d) Experimental characteristic reflection spectrum of one of the fabricated geometries. The dashed line is a Lorentzian fit to the experimental data.

**Mechanical response of the 1D-PhCW pillar cavity**

When light is coupled to an optical mode like the one reported in Figure 2d, the thermally activated mechanical oscillation modes of the 1D-PhCW pillar cavity can be optically transduced. This measurement relies on the dependence of the spectral position of the optical resonance with a pillar deformation, namely the optomechanical coupling, which leads to a modulation of the transmitted or reflected light that can be detected with large bandwidth near-IR photodetectors and processed with a spectrum analyser.[23] The radiofrequency (RF) spectrum evidencing the mechanical response of a representative 1D-PhCW pillar cavity of $\gamma=1$ and defect depth $g=0.75$ is reported in Figure 3a. This data is superimposed to the vacuum optomechanical coupling strength ($g_{OM}/2\pi$) simulated with a Finite-Element-Methods (FEM) solver plotted on the right axis.

The power spectral density (PSD) spectrum displays multiple mechanical modes appearing as narrow Lorentzian peaks. Within the 1D-PhCW pillar cavity, each individual pillar serves as a mechanical resonator anchored at one end (the substrate), vibrating at its own mechanical resonance frequency ($\Omega_{m,i}$, where the subindex i indicates a specific pillar of the cavity region). The mechanical signal exhibits a remarkable complexity, showcasing at least by four families of cantilever-like modes defined over specific spectral bands. Due to variations in the radii of the pillars forming the cavity, multiple peaks are observed within the spectral band covered by each family.

The related displacement field profile for the four families of cantilever-like mechanical modes is shown in the insets of Figure 3a. The motion associated with these modes and the corresponding displacement field profiles have been identified through FEM simulations. These families extend from the fundamental mode family, occurring at the lowest frequencies (<50MHz), up to the 4[th] harmonic mode family at higher frequencies (<900MHz). The computed values of $g_{OM}/2\pi$ are dominated by a moving interface contribution[24] despite the photoelastic one[25] (see Supplementary Section S4). Notably, the overall values exceed 1MHz for some of the mechanical modes within the first family of modes.

Even though the FEM simulations were performed using an idealized pillar profile (see Figure 1d) that slightly deviates from the fabricated one mostly on the undercut region, it is evident from Figure 3a that the computed frequencies for the four distinct mode families are in reasonably good agreement with the experimental observations. Interestingly, the signal amplitudes span over five orders of magnitude with the larger amplitudes belonging to the family of the fundamental modes and the weaker ones, corresponding to the fourth family of modes. This fact can be explained by the relative magnitudes of the $g_{OM}$ values calculated for each family, which are weighed by the frequency-dependent occupation of vibrational modes at the high temperature limit. The latter quantity is given by the average phonon number of the environment $n_{th} \approx k_B T_{bath}/\hbar\Omega_{m,i}$, where $T_{bath}$ is the ambient temperatures and $k_B$ and $\hbar$ the Boltzmann and Planck constants, respectively.[23]

Within each family, the pillars with smaller radii have smaller $\Omega_{m,i}$ compared to pillars with larger radii, i.e., the mechanical modes within the cavity region have $\Omega_{m,i}$ of decreasing values as the pillars are localized closer to the centre of the 1D-PhCW pillar cavity. Figure 3b illustrates this for the case of the fundamental family of mechanical modes by zooming at the low frequency region of Figure 3a. The experimental signal (bottom panel of Figure 3b) thus consists of a set of resonances, each corresponding to one pillar, with the frequency increasing for pillars located farther from the centre. To support the previous statement, in Figure 3c we show the deformation profile of modes situated at the two extremes of the fundamental family spectral band according to FEM simulations. The spatial distribution of the moving interface contribution to the $g_{OM}$ values is localized at the specific oscillating pillar, as expected (see Supplementary Section S4). In force sensing or biosensing applications, tracking the spectral shift of a mechanical RF peak could be directly related to perturbations at the location of the pillar providing that signal. This would enable extracting spatial maps of the perturbations, with a resolution determined by the spacing between adjacent pillars. These features are in sharp contrast to what displayed by suspended 1D-PhCW nanobeam cavities,[25–27] where the supported mechanical modes involve a collective oscillation of most of the cells composing the cavity.

The mechanical quality factors of the observed modes are on the order of $10^1$ for the first family of modes and $10^2$ for the rest of the families. Given the frequency range of the modes and the humidity of our lab (between 45 and 50%), mechanical losses are likely dominated by viscoelastic losses due to interaction with the surrounding medium[28].

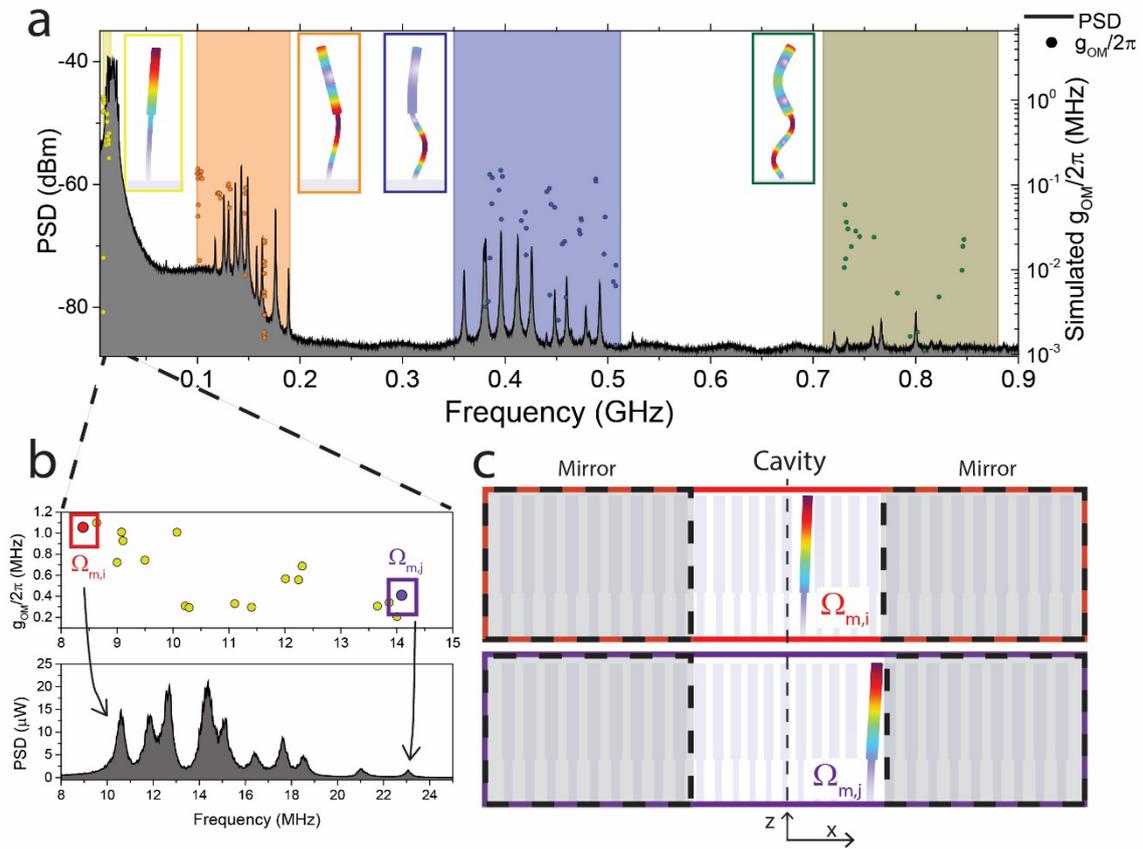

**Figure 3. Mechanical spectrum of the 1D-PhCW pillar cavity measured through optomechanical transduction.** a) RF spectrum, acquired by optomechanical measurement, evidencing the mechanical modes of vibration transduced by the 1D-PhCW pillar cavity. Four different families of cantilever-like modes, highlighted by colour boxes, can be identified from FEM simulations for the fundamental and up to the fourth harmonic. The related displacement field profile for each family is shown in the insets. The right axis and the scattered dots correspond to the vacuum OM coupling rate ($g_{OM}/2\pi$) calculations for a representative 1D-PhCW pillar cavity of $\gamma =1$ and defect depth g=0.75 with dimensions corresponding to the fabricated and measured geometry. b) Zoom of the fundamental family region. The experimental power spectral density (PSD) is plotted in linear scale in this case. Two mechanical modes placed at the extremes of the analysed spectral band are highlighted with coloured squares, whose deformation profile is depicted in panel c) using the same colour code.

To provide further insight into the effect of the geometrical parameters of the 1D-PhCW pillar cavities on their mechanical properties, we have made a systematic study of the behaviour of the mechanical spectra upon variation of the overall scaling factor $\gamma$ and the depth of the cavity region g (see Figure 4a and 4b respectively), keeping the other parameters fixed.

In the first case, varying the $\gamma$ factor (see Figure 4a), we observe that every family of mechanical modes (here only the first two are shown) increase its average frequency by increasing $\gamma$, which is indicated with the black arrows. This is an expected result that is consistent with the general behaviour of individual mechanical cantilevers, wherein a larger radius correlates with a higher oscillation frequency. It is worth noting that in this set of experiments we are also fixing the order of the optical mode under test, which in Figure 4a is the fundamental one.

When adjusting g while keeping $\gamma$ constant (see Figure 4b), we notice that the higher frequency side of the spectrum remains largely unchanged, whereas there is an expansion observed on the lower frequency side as g decreases. This is coherent with the fact that deeper defects are produced by smaller pillar radii near the centre of the cavity, resulting in lower natural frequencies in that region. Conversely, pillars located at the cavity periphery exhibit minimal variations in response to modifications of g. We have focused the set of experiments performed in Figure 4b on the fundamental family of mechanical modes and used the second order optical mode. The behaviour of the mechanical frequencies with g is indicated with black arrows.

Finally, in Figure 4c, we have fixed the geometrical parameters of the 1D-PhCW pillar cavity and explore the variations of the mechanical spectrum, focused on the fundamental family, as the different supported optical cavity modes are excited. In this case, given that we are dealing with the same geometry, we observe that the set of transduced mechanical modes remains consistent across the panels, as denoted by the dashed vertical lines indicating their spectral position. However, the relative signal strength associated with each mechanical peak vary among the figure panels. This variability directly stems from the different electromagnetic field spatial distribution along the 1D-PhCW pillar cavity for the distinct supported optical modes, an observation that is also reinforced with FEM simulations (see Supplementary Section S4).

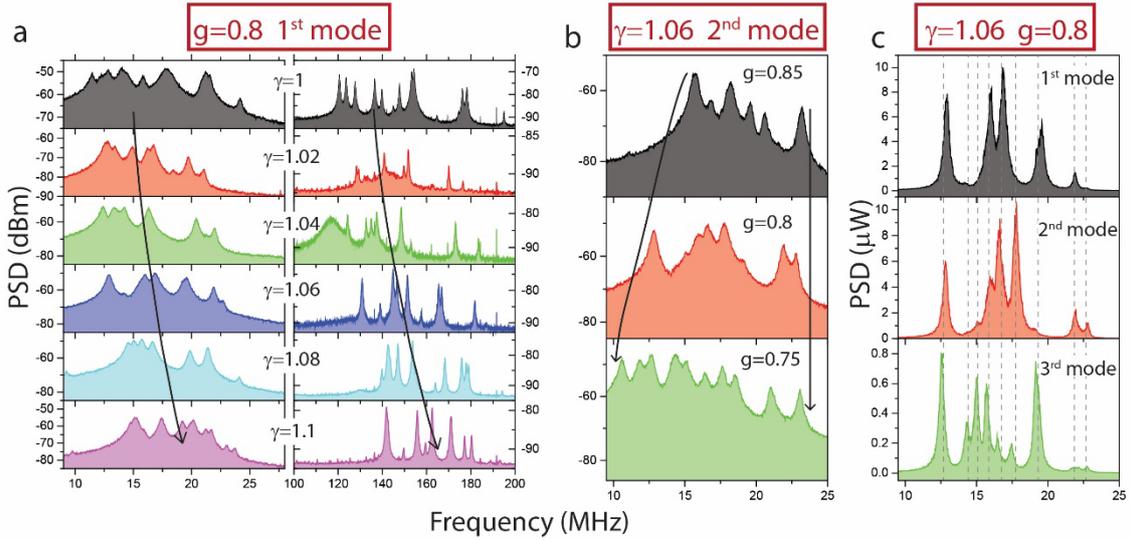

**Figure 4. Mechanical spectrum dependence of the geometrical parameters.** a) RF spectrum for different values of the scaling factor γ for the fundamental and second order family of mechanical modes (left and right, respectively). The defect depth is g=0.8 and the optical mode under test is the fundamental one. Black arrows indicate the general tendency of the set of observed peaks. b) RF spectrum for different values of the defect depth g for the fundamental family of mechanical modes. The scaling factor is γ=1.06 and the optical mode under test is the second order one. Black arrows indicate the general tendency at the extremes of the spectrum. c) RF spectrum for a fixed geometry and different optical modes under test. The scaling factor is γ=1.06 and the defect depth is g=0.8. Black dashed lines indicate the spectral position of the mechanical modes. PSD is reported in linear scale in this case.

To illustrate the sensitivity of the 1D-PhCW pillar cavities to physical perturbations, we brought the tapered fibre used for optical probing into contact with the pillars. We recorded the evolution of the RF spectrum, focusing on the spectral band covered by the second family of mechanical modes, while manually dragging the fibre back and forth along the x-direction over the tops of the pillars (see Figure 5). Interestingly, several RF peaks were perturbed and significantly shifted in spectral position during the measurement. We attribute the observed linewidth broadening of these perturbed RF peaks to damping of the mechanical modes caused by physical contact with the fibre. Supporting our interpretation, some of the broad peaks around 180 MHz shift as the fibre is dragged and then suddenly become narrow and stable. We associate this behaviour with the moment the fibre stops touching the specific pillars generating those RF peaks. Figure 5a shows three RF spectra recorded for different positions of the fibre over the 1D-PhCW pillar cavity, corresponding to the moments highlighted in Figure 5b with white dashed lines. The specifics of the interaction between the fibre and the pillars leading to the observations of Figure 5 are beyond the scope of this manuscript.

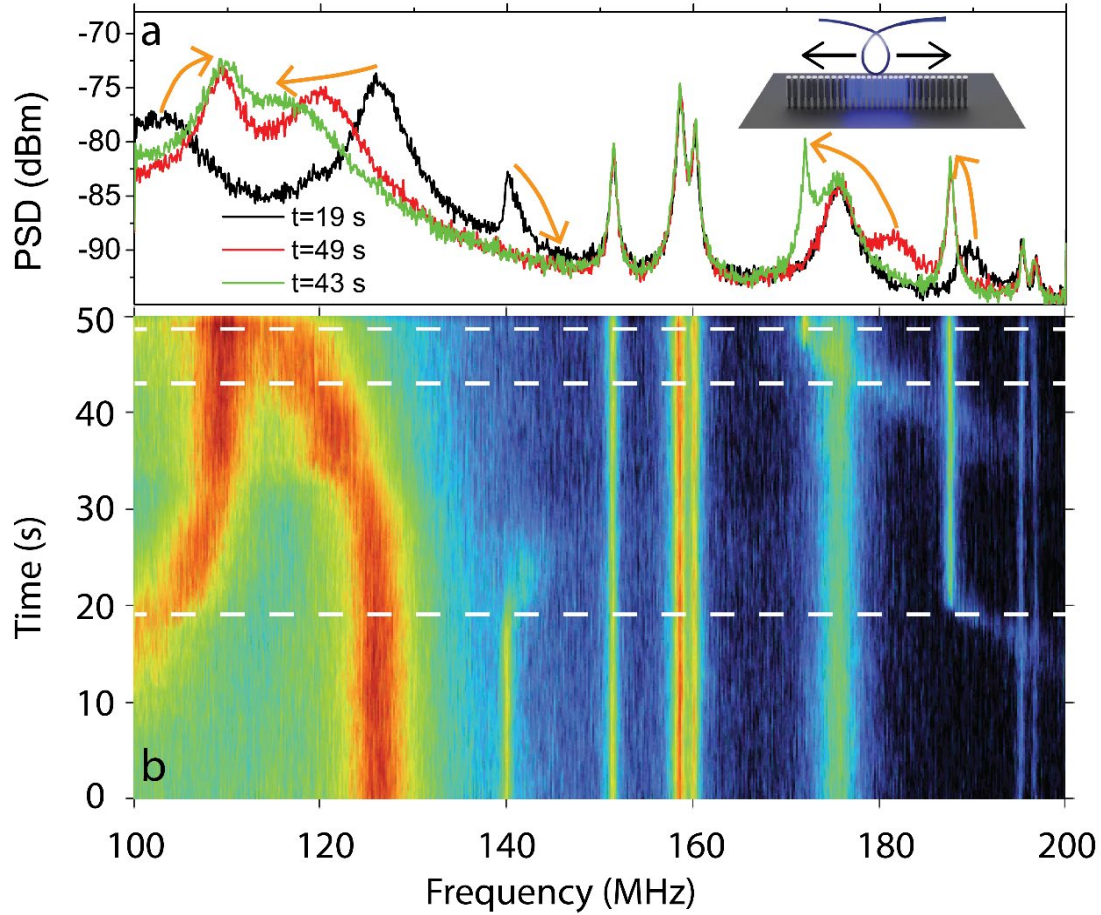

**Figure 5. Evolution of the mechanical spectrum with the positioning of the tapered fibre on top of the 1D-PhCW pillar cavity while in physical contact.** a) RF spectrum for three different positions of the fibre, with arrows indicating the evolution of the peaks across the measurements. The inset shows a sketch of the relative positioning between the fibre and the geometry. b) Temporal evolution of the RF spectrum as the fibre is manually dragged back and forth over the tops of the pillars. The dashed lines indicate the moments at which the RF spectra displayed in panel a) were taken.

Conclusions

In this work we have introduced a novel platform for integrated photonics, featuring a fully silicon-based 1D photonic crystal waveguide (1D-PhCW) composed of nanopillars. By carefully engineering the dimensions of the silicon pillars, we achieved vertical light confinement and established an energy bandgap in the near-infrared spectrum for TM polarization. Through experimental demonstrations, we established optical cavities with quality factors exceeding $10^3$ by incorporating defects within the periodic 1D-arrays of pillars. We acknowledge challenges associated with optical scattering losses likely associated to the surface rough of the undercut portion of the pillars. However, our comprehensive understanding of the mechanical and optical properties of these 1D-PhCW pillar cavities opens avenues for mitigating this issue and further optimizing their optical performance.

A significant advantage of our approach is the dual functionality of the nanopillars, which naturally behave as mechanical cantilevers. This characteristic enables our fabricated structures to serve as excellent optomechanical photonic crystal cavities, facilitating the optical transduction of mechanical motion. Remarkably, our geometries exhibit optomechanical (OM) coupling rates large enough to enable the detection of thermally activated motion up to the 4$^{th}$ order of cantilever-like modes. These combined features could, for instance, enable non-invasive, label-free, means of sensing variations of the pillar mechanical properties upon forces exerted by, for instance, biological specimens deposited on the surface of the 1D-PhCW pillar cavities. Furthermore, since there exists a direct correlation between the oscillation frequency of a pillar and its spatial position within the cavity, achieving spatial resolutions on the order of the pitch, i.e., a few hundred nanometers, would be achievable.

Our design offers several advantages, including enhanced mechanical properties, cost-effectiveness, integration possibilities, and scalability. Notably, our structures present a new path in front to conventional suspended Si beam OM cavities fabricated on silicon-on-insulator (SOI) substrates.

In summary, our work represents a significant step towards realizing practical photonic devices based on 1D-PhCW composed of nanopillars with enhanced functionalities, offering opportunities applications in force sensing, biosensing, and beyond.

**Acknowledgements**: This work has been financially supported by the European Union's H2020 FET-OPEN project STRETCHBIO (GA 964808).

## S1. Experimental setup details

The experiments were made in a standard set-up to characterize optical and mechanical properties of optomechanical cavities illustrated in Figure 2a of the main text. To cover the wavelength range from 1355 nm up to 1640 nm, two lasers were used: a Santec TSL-570 for the range between 1355 nm to 1490 nm, and a Yenista Tunics T100S for the range from 1440 nm up to 1640 nm. The tapered silica fiber is connected to these lasers, and it first passes through a fiber polarization controller (FPC) and then a circulator. The FPC enables us to adjust the polarization of the input light to optimize coupling efficiency and minimize polarization-dependent losses. The circulator enables light to travel through each port in only one direction, and it is used to separate the transmitted and reflected light. Each branch is connected to a photodetector, which measures the power of the transmitted/reflected light at each wavelength.

To position the fiber loop on the cavity region we use a 50x microscope objective and a CCD camera to capture zenital images. The positioning is manually adjusted with a submicrometer precision positioning system.

To check for the presence of a radiofrequency (RF) modulation of the optical signal we connect the output of the photodetectors to the 50 Ω input impedance of a spectrum analyser (SA) with a bandwidth of 13.5 GHz. All the measurements were performed in an anti-vibration cage at ambient conditions.

## S2. Fabrication details

The one-dimensional photonic crystal waveguides (1D-PhCW) pillar structures were fabricated by electron beam (e-beam) lithography and reactive ion etching (RIE) on an n-type <100> silicon substrate from Siegert Wafer (Germany) with a specified resistivity of 1-20 Ohm-cm (refer to Figure S1).

The wafer was prepared for electron beam exposure by spin-coating a 180 nm layer of CSAR positive e-beam resist (AR-P 6200, Allresist GmbH, Germany). The desired pattern was exposed using a JEOL JBX-9500FS electron beam lithography system with a dose of 300 μC/cm2 and subsequently developed using developer AR 600-546 (Allresist GmbH, Germany) for 60 seconds. A 20 nm Aluminium layer was deposited using e-beam evaporation followed by a lift-off process (Microposit Remover 1165) to remove the remaining CSAR resist, leaving the aluminium layer with the desired pattern as a mask for the subsequent dry etch.

The RIE process was done in 2 steps to create the upper and lower sections of the nanopillar with different diameters. The first part of the etching process was done with a simultaneous mix of etching gas (SF6 at 44 sccm) and passivation gas (C4F8 at 77 sccm) for 4 minutes with a coil power of 1000W and a platen power of 20W. For the second part of the process, the etch was changed to a cyclic process alternating between etching with SF6 and passivation with C4F8 for 25 cycles. The height of the nanopillars can be adjusting by changing the etching time in the first step and the number of cycles in the second step. Finally, the remaining aluminium mask was removed using a TMAH based developer (AZ 726 MIF, Microchemicals GmbH, Germany).

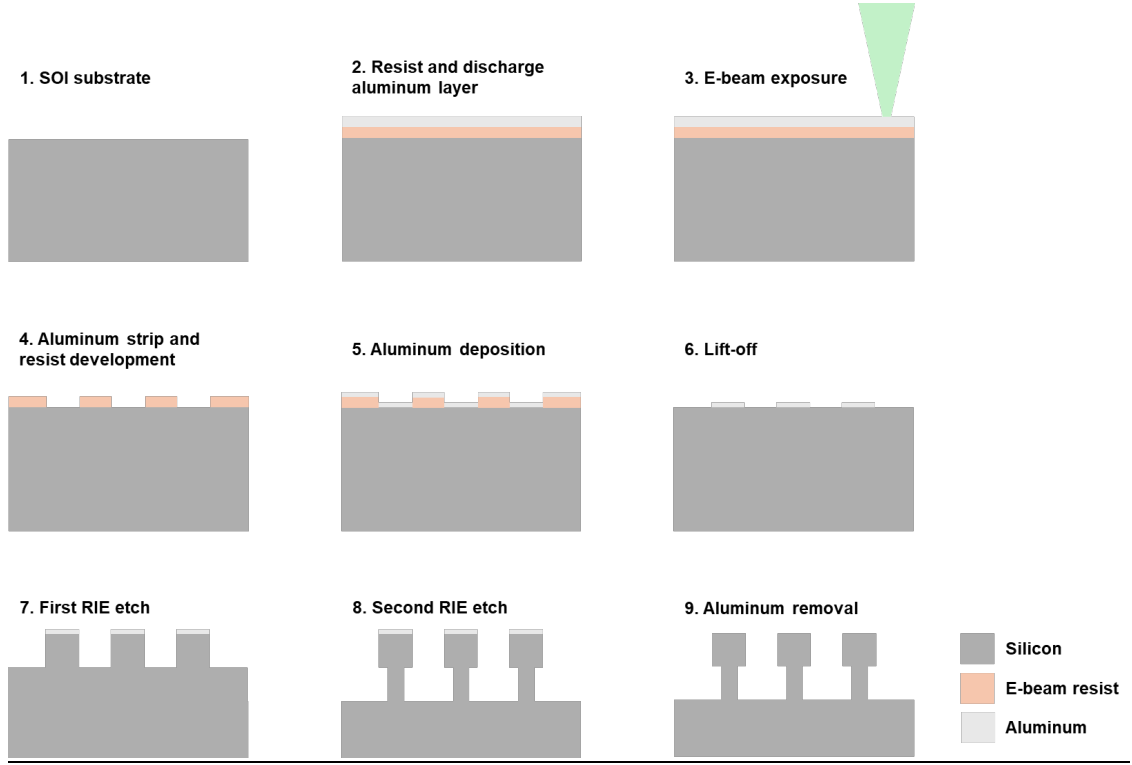

**Figure S1.** Sketch of the fabrication steps required to fabricate the 1D-PhCW pillar structures.

## S3. Optimization of the optical quality factor by design

In this section we discuss on the geometrical requirements to generate a high quality-factor optical state within a 1D-PhCW made of silicon pillars. We use the optical quality-factor as a figure of merit for the optical cavities, which has been evaluated for the fundamental optical mode by calculating the ratio between the imaginary and the real parts of the optical eigenfrequencies.

The cavity region has been created as described in the main text. It consists of two mirror regions, each one of them constructed by a repetition of the same unitary cell about 9 times. Within them, we have introduced 11-12 cells with a gradual decrease of the pitch and radius of the pillars towards the centre. With this geometrical configuration of the optical cavity, the simplest cross section that can be simulated is that with the pillars surrounded by air, which is the case illustrated in Figure S2a. With this cross section it is possible to create cavity modes with radiative optical quality factors exceeding $10^5$ at about 200 THz for pillar heights of $t_1$=1500nm. If the lower substrate is replaced by a solid one, such as SiO$_2$ as illustrated in Figure S2b, there appears a significant light leakage towards the substrate, limiting the quality factors of the cavity mode to below 10$^3$. We believe that this is the main reason explaining why there are no previous experimental realisations of 1D-PhCW made of silicon pillars.

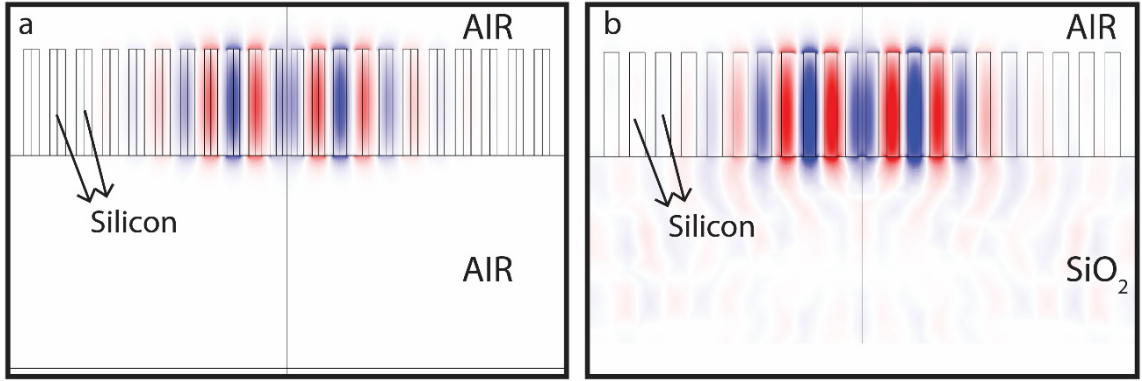

**Figure S2.** **Optical modes of 1D-PhCW pillar cavities.** Finite-Element-Method simulation of the electric field along the z direction (E$_z$) of the TM fundamental optical cavity mode as seen from the side. a) The pillar array is surrounded by air. b) The pillar array leans on top of a SiO$_2$ substrate.

The previous results indicate that a geometrical innovation such as the one proposed by us in the main text must be introduced to obtain a high-quality 1D-PhCW made of silicon pillars deposited on a substrate.

In the following studies we have fixed the silicon pillar geometry to that consisting of two vertically stacked silicon pillars, i.e., a top pillar of radius r with a height $t_1$ resting upon another Si pillar of radius r-Δr with a height $t_2$. The radii and pitch values have been fixed values to r=105nm, Δr=50nm and pitch a=350nm.

Figure S3 reports the evolution of the Q-factor as a function of the height of the lower pillar $t_2$, while $t_1$ has been fixed to $t_1$=1500nm. The optical Q-factor increases dramatically by improving the isolation from the substrate, i.e., by increasing $t_2$, while the mode frequency does not significantly shift (not plotted). If $t_2$<200nm the light leakage towards the substrate dominates, forbidding any supported optical mode confined on the top part of the pillar. The height of the fabricated structures ($t_2$~850nm) would ensure radiative optical Q-factor on the order of 5x10$^4$, which is more than an order of magnitude greater than what measured experimentally. Therefore, the observed optical losses, which lead to experimental Q-factors of 10$^3$, probably stem from surface scattering losses at the lower part of the pillars.

It is worth noticing that, for $t_2$>200nm, FEM simulations indicate that the differences between using a Silicon or a SiO$_2$ as substrate are not significant. There are still supported modes in the case of using $SiO_2$ as substrate if $t_2<200$nm, but the Q-factors lie below $10^3$. For a value of $t_2=0$, the results are those of Figure S1b.

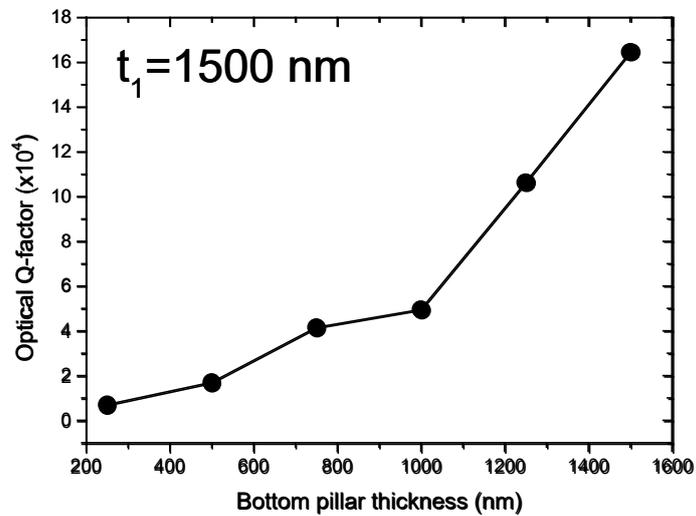

**Figure S3.** **Evolution of the optical Q-factor with the height of the bottom part of the pillar.** The height of the upper part of the pillar is fixed to $t_1=1500$nm

Finally, we have studied the effect of modifying the top pillar thickness $t_1$ while keeping $t_2=1500$nm. Figure S4 shows that, in this case, both the Q-factor (black curve, left axis) and the spectral position of the cavity mode (red curve, right axis) are affected, in a way that the former improves with $t_1$ while the latter shifts towards smaller frequencies. The cut-off for supporting an optical mode is found at $t_1=500$nm.

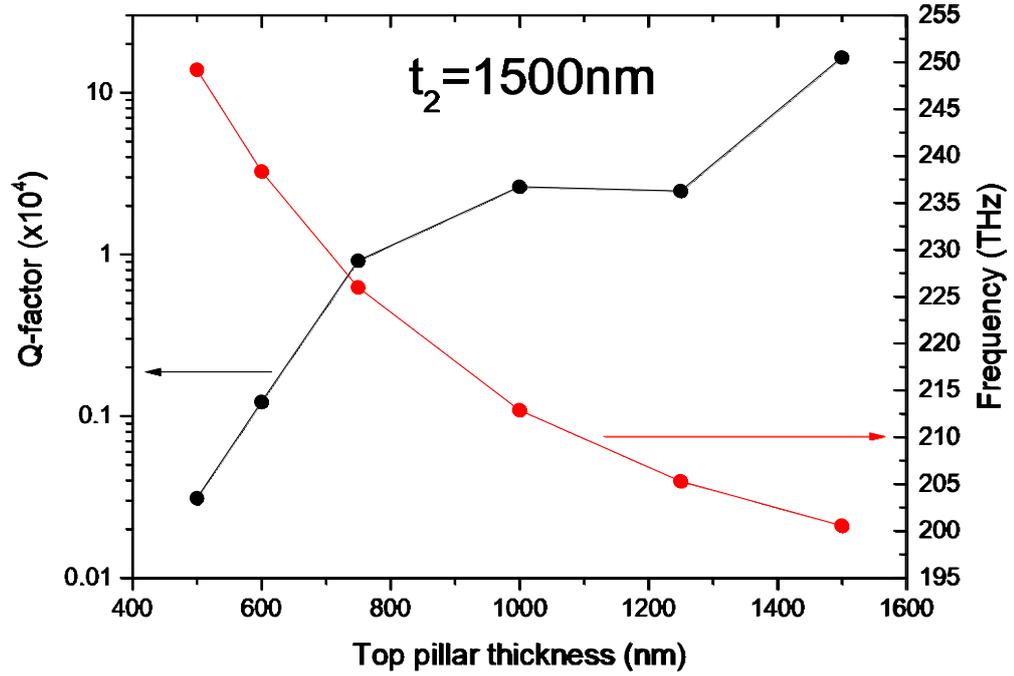

**Figure S4.** **Evolution of the optical Q-factor (black curve, left axis) and spectral position of the optical mode (red curve, right axis) with the height of the upper part of the pillar.** The height of the bottom part of the pillar is fixed at $t_2$=1500nm

## S4. OM coupling calculations.

Single-particle optomechanical coupling rates ($g_{OM}$) between optical and mechanical modes are estimated by considering both photo-elastic (PE) and moving-interface (MI) effects[1,2]. The PE effect is a result of the acoustic strain within bulk silicon while the MI mechanism comes from the dielectric permittivity variation at the boundaries associated with the deformation.

The calculation of the MI coupling coefficient $g_{MI}$ is performed using the integral given by Johnson et al.[1]:

$$g_{MI} = -\frac{\pi \lambda_r}{c} \frac{\oint (\mathbf{Q} \cdot \hat{\mathbf{n}})(\Delta\varepsilon E_{\parallel}^2 - \Delta\varepsilon^{-1} D_{\perp}^2) dS}{\int \mathbf{E} \cdot \mathbf{D} dV} \sqrt{\hbar/2m_{eff}\Omega_m}$$

(S1)

where **Q** is the normalized displacement (max{|**Q**|}=1), **n̂** is the normal at the boundary (pointing outward), **E** is the electric field and **D** the electric displacement field. $\varepsilon$ is the dielectric permittivity, $\Delta\varepsilon = \varepsilon_{silicon} - \varepsilon_{air}$, $\Delta\varepsilon^{-1} = \varepsilon^{-1}_{silicon} - \varepsilon^{-1}_{air}$. $\lambda_r$ is the optical resonance wavelength, $c$ is the speed of light in vacuum, $\hbar$ is the reduced Planck constant, $m_{eff}$ is the effective mass of the

mechanical mode and $\Omega_m$ is the mechanical mode eigenfrequency, so that $\sqrt{\hbar/2m_{eff}\Omega_m}$ is the zero-point motion of the resonator.

A similar result can be derived for the PE contribution[2]:

$$g_{PE} = -\frac{\pi \lambda_r}{c} \frac{\langle E|\delta\varepsilon|E\rangle}{\int \boldsymbol{E}\cdot\boldsymbol{D}dV} \sqrt{\hbar/2m_{eff}\Omega_m} \qquad (S2)$$

where $\delta\varepsilon_{ij} = \varepsilon_{air} n^4 p_{ijkl} S_{kl}$, being $p_{ijkl}$ the PE tensor components, $n$ the refractive index of silicon, and $S_{kl}$ the strain tensor components.

The addition of both contributions results in the overall single-particle OM coupling rate. However, we have observed that the PE contribution to $g_{OM}$ is negligible with respect to the MI one and thus we can assume that $g_{OM} \sim g_{MI}$.

To provide further insight on the optomechanical properties of the 1D-PhCW cavity made of silicon pillars, in Figure S5 we represent the spatial contributions relevant for the calculation of $g_{MI}$, i.e., the integrand of Eq. S1. We have focused on the fundamental optical mode (refer to Figure S5a) and the mechanical modes highlighted in Figure 3 of the main text, which are associated to a pillar close to the centre ($\Omega_{m,i}$=8.39MHz, refer to Figure S5b) and a pillar at the side of the cavity region ($\Omega_{m,j}$=14.09MHz, refer to Figure S5c), respectively. As expected, the contribution to $g_{MI}$ is just confined to the oscillating pillar (refer to Figure S5d and Figure S5e), which implies that the resonant peaks that are measured in the RF spectra of the optical signal are associated to the mechanical oscillation of single pillars placed within the cavity region.

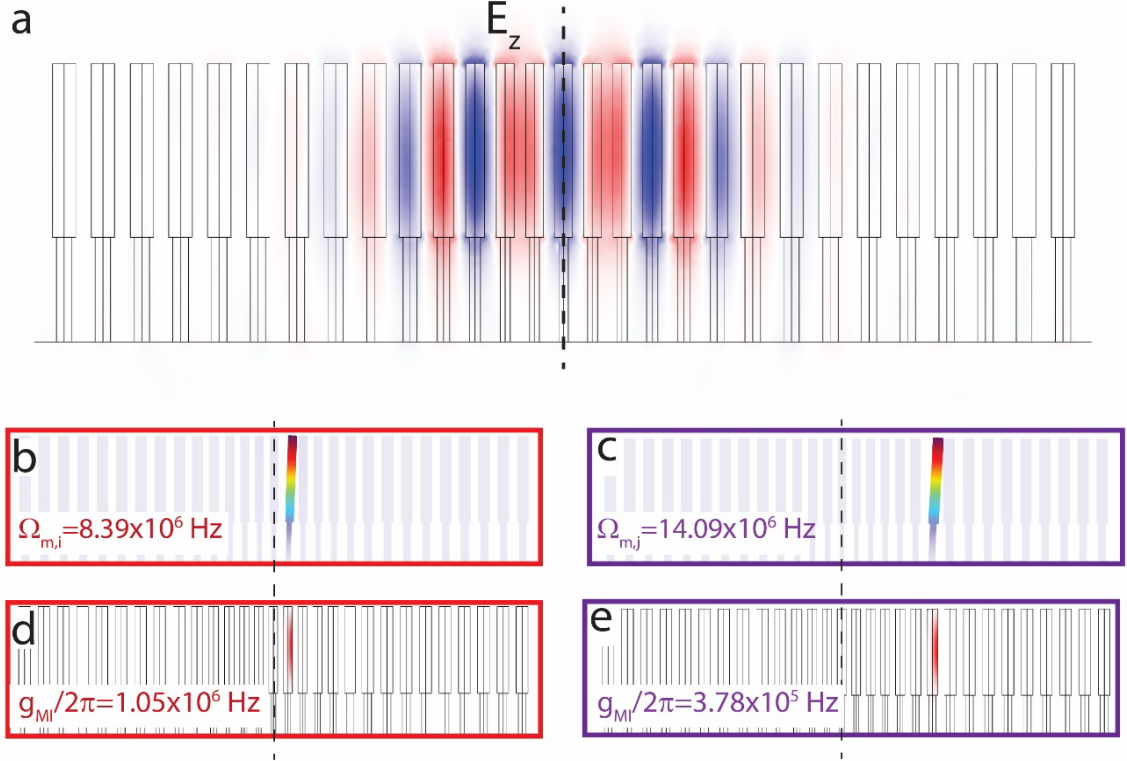

**Figure S5.** **Optomechanical coupling between the fundamental optical mode and mechanical modes of the pillars belonging to the cavity region** a) Finite-Element-Method simulation of the electric field along the z direction ($E_z$) of the TM fundamental optical cavity mode as seen from the side. b) and c) Normalized mechanical displacement field |Q| of mechanical modes at 8.39 MHz (panel b) and 14.09 MHz (panel c). d) and e) Normalized surface density of the integrand in Eq. S1, showing the contributions to $g_{MI}$ associated to the modes of panels b and c, respectively.

It is worth noting that, outside the cavity region, the simulated mechanical modes are collective oscillations of several pillars of the mirror region (not shown), which is a consequence of their identical geometry and the slight mechanical coupling through the substrate. Those mechanical modes do not exhibit significant $g_{OM}/2\pi$ values given that the overlap with the optical mode is modest.

Finally, we discuss the role of the spatial distribution of the optical mode on tailoring the RF spectra that we reported experimentally in Figure 4c of the main text. There we show that the relative signal strength associated with each mechanical peak vary with the optical mode used to excite the cavity and associate that observation to a different electromagnetic field spatial distribution along the 1D-PhCW pillar cavity. Figure S6a and Figure S6b report that the spatial distribution of the fundamental and second order optical mode extracted from the FEM simulations are clearly different, showing that the second order mode occupies more volume than the fundamental, even extending to pillars of the mirror region. The values of $g_{OM}/2\pi$ associated with the mechanical modes of the first family are therefore different for each of the optical modes considered. Given that the mechanical frequencies increase with the distance of the oscillating pillars from the centre, the frequency spectrum of $g_{OM}/2\pi$ roughly follows the spatial distribution of the optical field in the cavity region.

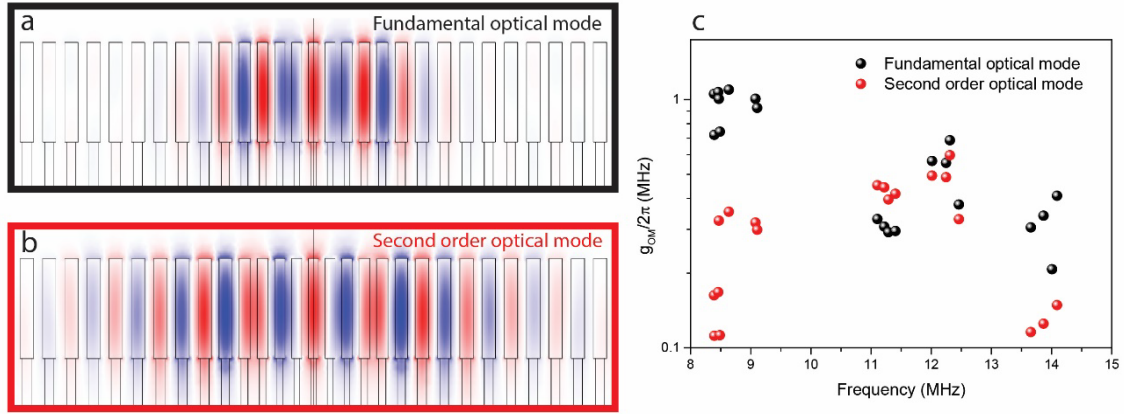

**Figure S6.** a) and b) Finite-Element-Method simulation of the electric field along the z direction ($E_z$) of the TM fundamental and second order optical cavity mode (panel a and b, respectively) as seen from the side. c) Vacuum OM coupling ($g_{OM}/2\pi$) calculations for the fundamental and second order optical cavity modes supported by the cavity (black and red dots, respectively). The mechanical modes are the same for both calculations.